\documentclass[conference,a4paper]{IEEEtran}
\IEEEoverridecommandlockouts
\usepackage{float}
\usepackage{amsmath}
\usepackage{amsthm}
\usepackage{amssymb}	
\usepackage{graphicx}
\usepackage{subfigure}
\usepackage{enumitem}
\usepackage{algorithm}
\usepackage{stfloats}
\usepackage{hhline}
\usepackage{cite}
\usepackage{cuted}
\usepackage{color}
\newcommand*{\J}{\jmath}%

\usepackage{amsmath,amssymb,mathtools,bm,etoolbox}
\usepackage[center]{caption}

\DeclarePairedDelimiterXPP\Aver[1]{\mathbb{E}}{[}{]}{}{
	
	#1
}

\newtheorem{my_theorem}{Theorem}

\title{RIS-Assisted Vehicular Network with Direct Transmission over Double-Generalized Gamma Fading Channels}

\author{
	\IEEEauthorblockN{ Vinay Kumar Chapala, Arsalan Malik, and S.~M.~Zafaruddin}\\
	\IEEEauthorblockA{ Department of Electrical and Electronics Engineering, 
		BITS Pilani, Pilani Campus, Pilani-333031, Rajasthan, India
		\\ Email: \{p20200110, f20190499, syed.zafaruddin\}@pilani.bits-pilani.ac.in}
	
	\thanks{This work was supported in part by the Science and
		Engineering Research Board (SERB), Department of Science and Technology
		(DST), Government of India, under MATRICS Grant MTR/2021/000890 and Start-up Research Grant SRG/2019/002345.}
}

\begin{document}
	\maketitle 
	\begin{abstract}
	
		Reconfigurable intelligent surface (RIS) can provide  stable connectivity for vehicular communications when direct transmission  becomes significantly weaker with dynamic channel conditions between an access point and a moving vehicle. In this paper,  we analyze the performance of a RIS-assisted vehicular network by coherently combining received  signals  reflected by RIS elements and direct transmissions from the source terminal over double generalized Gamma (dGG) fading channels. We present  analytical expressions on the outage probability and average {\color{black}bit-error rate} (BER) performance of the considered system by deriving exact density and distribution functions for the end-to-end signal-to-noise ratio (SNR) resulted from the finite sum of the direct link and product of channel coefficients each distributed according to the dGG. We also develop asymptotic analysis on the outage probability and average BER to derive diversity order for a better insight into the system performance at high SNR. We validate the derived analytical expressions through numerical and simulation results and demonstrate  scaling of the system performance with RIS elements and a comparison to the conventional relaying techniques and direct transmissions considering various practically relevant scenarios.
			
	\end{abstract}				
	
	\section{Introduction}
The upcoming 6G wireless system envisions to cater exceedingly higher requirements of network throughput,  lower latency of transmission, and stable  connectivity  for autonomous vehicular communications \cite{Yishi_2021}.  However, dynamic channel conditions between an access point to a  vehicle in motion may be a bottleneck for the desired quality of service. The use of 	reconfigurable intelligent surface  (RIS) can be a promising technology to create a strong line-of-sight (LOS) transmissions by artificially controlling the characteristics of propagating signals for vehicular communications \cite{Yishi_2021, Qingqing2021,Basar2019_access}. An RIS is a planar metasurface made up of a large number of inexpensive passive reflective elements that can be electronically programmed to customize the phase of the incoming electromagnetic wave for enhanced signal quality and transmission coverage. As is for other wireless networks, it is desirable to analyze the applicability  of RIS for  vehicular communications as a proof of concept considering generalized deployment scenarios.
	
	Recently, there has been an extensive study exploring the benefits of the RIS approach for  variety of wireless communication networks such as radio-frequency (RF)  \cite{Kudathanthirige2020, Qin2020, Ferreira2020,   Selimis2021, Khoshafa,trigui2020_fox, du2021}, free-space optics (FSO)   \cite{Jamali2021, chapala2021unified},  and terahertz (THz) \cite{du2020_thz, chapala2021THz}. More specifically, the authors analyzed the performance of RIS-assisted RF systems over Rayleigh fading model in \cite{Kudathanthirige2020}, Rician fading in \cite{Qin2020}, Nakagami-m channel model in \cite{Ferreira2020,  Selimis2021} and fluctuating two rays (FTR) fading model for RIS-assisted mmWave communications in \cite{du2021}. The authors in \cite{trigui2020_fox} presented an exact performance analysis for RIS-assisted wireless transmissions over generalized Fox’s H fading channels. In \cite{chapala2021unified}, an  unified  performance analysis for a FSO system was presented considering different atmospheric turbulence models and pointing errors. The authors in \cite{du2020_thz, chapala2021THz}  analyzed  RIS-aided THz communications over different fading channel models combined with antenna misalignment and hardware impairments.

Considering the benefits for conventional wireless transmissions, 	the use of RIS has been recently proposed for connected autonomous vehicles and vehicular networks \cite{Wang_2020_outage,Kehinde_2020,dampahalage2020intelligent,Long_2021, Ozcan2021,Abubakar_2020}. The authors in \cite{Wang_2020_outage} used series expansion and the central limit theorem (CLT) to approximate the outage probability of RIS-assisted vehicular networks over Rayleigh and Rician fading channels. In \cite{Kehinde_2020}, the authors analyzed the outage probability, average {\color{black}bit-error rate} (BER), and ergodic  capacity of a RIS-enabled mobile network with random user mobility. The authors \cite{dampahalage2020intelligent} developed an iterative algorithm to optimize the RIS phase shifts for the data rate maximization  of a RIS-assisted mmWave vehicular communication network. In \cite{Long_2021}, the authors studied the RIS-enabled vehicle-to-vehicle (V2V) communications using the Fox's H-function distribution. In \cite{Ozcan2021}, {\color{black}the  placement of multiple RIS} is optimized for vehicle-to-everything (V2X) communications. In \cite{chapala2021multiris}, we analyzed the performance of multiple RIS-based vehicular communications over  double Generalized-Gamma (dGG) fading model. In the aforementioned research,  the direct signal received from the source is ignored.	It should be noted that there can be a direct transmission link from the  source to the destination in addition to the reflected signal through RIS elements, which has been studied in  \cite{Selimis2021,Khoshafa} \cite{dampahalage2020intelligent,Ozcan2021}.  {\color{black}However, the performance analysis is  limited to asymptotic bounds and   approximations considering simpler fading models. To the best of authors' knowledge an exact analysis for RIS-aided transmissions with direct link over double generalized (dGG) fading  that accurately models the vehicular communications has not been reported in the literature \cite{Petros2018}. It should be mentioned that deriving  the statistics of the end-to-end SNR for the finite sum of the direct link and product of channel coefficients each distributed according to the dGG is quite involved.}

 In this paper, we analyze the performance of a RIS-assisted transmission scheme  by coherently combining received  signals  reflected by RIS elements and direct transmissions from the source terminal for an enhanced connectivity in vehicular communications. We present  analytical expressions on the outage probability and average BER performance  by deriving exact density and distribution functions for the end-to-end SNR of the considered system. We also develop diversity order of the system by deriving   asymptotic expressions on the outage probability and average BER for  better Engineering insights on the system performance at high SNR. We  demonstrate  scaling of the system performance with RIS elements and a  comparison to the conventional relaying techniques and direct transmissions considering various practically relevant scenarios.

{\color{black}	

\emph{Notations}:   {\footnotesize{ $\Gamma(t) =\int_{0}^{\infty}s^{t-1}e^{-s}ds$}}  denotes the Gamma function, {\footnotesize{ $\Gamma(a,t) =\int_{t}^{\infty}s^{a-1}e^{-s}ds$}}  denotes the incomplete Gamma function,  {\footnotesize{$H_{p,q}^{m,n}  \Big(\begin{array}{c}x\end{array} \Big\vert \begin{array}{c} (a_1,A_1), \ldots (a_p,A_p)\\ (b_1,B_1), \ldots (b_q,B_q) \end{array}\Big)$}} denotes the Fox's H function, a notation for $\{a_{i}\}_{1}^{N} = \{a_{1},\cdots,a_{N}\}$, and $\J$ denotes the imaginary number.
}
	\section{System Model}\label{sec:system_model}
	
\begin{figure}
		\centering
		\vspace{-26mm}
		\includegraphics[scale=0.35]{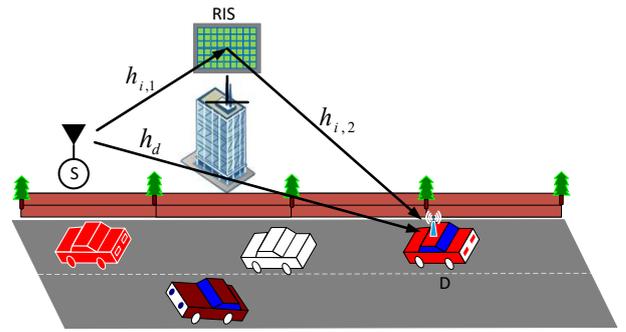} \vspace{-6mm}
		\caption{System Model.}\vspace{-4mm}
		\label{fig:system_model}
	\end{figure}

	We consider a  transmission model where a {\color{black}single-antenna} source ($S$) communicates to {\color{black}a single-antenna} destination ($D$) through an RIS with $N$ reflecting elements, as shown in Fig.~\ref{fig:system_model}. We also consider direct transmission link which may be useful for an enhanced vehicular connectivity.  We consider that the  elements of RIS are  spaced half of the wavelength and assume independent channels at the RIS \cite{Basar2019_access, du2021}. Assuming perfect  knowledge of  channel phase at each RIS element, the signal received at the destination is given as \cite{Selimis2021}:
	\begin{eqnarray}\label{model_smpl}
		y = \sqrt[]{P_{t}} s \big( h_{l_{ris}} \sum_{i=1}^{N} \lvert h_{i,1} \rvert \lvert h_{i,2} \rvert + h_{l} \lvert h_d \rvert\big) + v
	\end{eqnarray}
	where $P_{t}$ is the transmit power, $s$ is the unit {\color{black}power} information bearing signal, $|h_{i,1}|$  and  $|h_{i,2}|$  are channel fading coefficients between the source to the $i$-th RIS element and between the $i$-th RIS element to the destination, respectively, $\lvert h_d\rvert$ is the flat fading coefficient between source and destination, and $v$ is the additive Gaussian noise with zero mean and variance $\sigma_v^2$.	
		
		The path loss $h_{l_{ris}}$ of the cascaded link can be modeled using the free-space channel modeling of RIS-assisted communications \cite{Tang_2021, du2020_thz} as $h_{l_{ris}}=\frac{\sqrt{G_{t}G_{r}}c^{2}}{16\pi f^{2}d_{1}d_{2}}$, where $G_{t}$, $G_{r}$ represent transmit and receive antenna gains, $c$ is speed of the light, $f$ is frequency of operation, $d_{1}$ and $d_{2}$ are the distances from source to RIS and RIS to destination respectively. The path loss component of direct transmission is given as $h_{l}=\frac{\sqrt{G_{t}G_{r}}c}{4\pi f {\color{black}\sqrt{d_{1}^{2}+d_{2}^{2}}}}$. {\color{black}We ignore the heights of source	and destination to compute the path loss for direct link.}	We assume short-term fading coefficients $|h_{i,1}|$, $|h_{i,2}|$, and $|h_d|$ to be independent but non-identical distributed according to the  dGG \cite{Petros2018}. The Fox's H representation of the  {\color{black}probability density function} (PDF) for a dGG random variable $X(\alpha_1, \beta_1, \alpha_2, \beta_2)$ is given by \cite{chapala2021multiris}:
	\begin{eqnarray}\label{eq:dgg_pdf}
		f_{X}(x) = \psi x^{\alpha_{2}\beta_{2}-1} H_{0,2}^{2,0} \Bigg[\begin{array}{c} \phi x^{\alpha_{2}} \end{array} \big\vert \begin{array}{c} - \\ (0,1),(\frac{\alpha_{1}\beta_{1}-\alpha_{2}\beta_{2}}{\alpha_{1}},\frac{\alpha_{2}}{\alpha_{1}})\end{array}\Bigg]
	\end{eqnarray}
	where $\psi=\frac{\alpha_{2}}{(\frac{\Omega_{1}}{\beta_{1}})^{\frac{\alpha_{2}\beta_{2}}{\alpha_{1}}}(\frac{\Omega_{2}}{\beta_{2}})^{\beta_{2}}\Gamma(\beta_{1})\Gamma(\beta_{2})}$, $\phi=\frac{\beta_{2}}{\Omega_{2}} \big(\frac{\beta_{1}}{\Omega_{1}}\big)^{\frac{\alpha_{2}}{\alpha_{1}}}$,  $\alpha_{1}$, $\beta_{1}$, $\alpha_{2}$, $\beta_{2}$ are Gamma distribution shaping parameters and $\Omega_{i}=\big(\frac{\mathbb{E}[\chi_{i}^{2}]\Gamma(\beta_{i})}{\Gamma(\beta_{i}+2/\alpha_{i})}\big)^{\alpha_{i}/2}$, $i=1,2$ is the $\alpha_{i}$-root mean value.

	\section{Performance Analysis}
	To facilitate performance analysis, the distribution function of  $Z=h_{\rm RIS} + h_d=\sum_{i=1}^{N} \lvert h_{i,1} \rvert \lvert h_{i,2} \lvert + \lvert h_d \rvert $ is required, where  $\lvert h_{i,1} \rvert\thicksim d\mathcal{GG}(\alpha_{i,1},\beta_{i,1},\alpha_{i,2},\beta_{i,2})$, $\rvert \lvert h_{i,2} \lvert\thicksim d\mathcal{GG}(\alpha_{i,3},\beta_{i,3},\alpha_{i,4},\beta_{i,4})$ and $\lvert h_d \rvert\thicksim d\mathcal{GG}(\alpha_{d,1},\beta_{d,1}, \alpha_{d,2},\beta_{d,2})$. We denote $\{a_{i}\}_{1}^{N} = \{a_{1},\cdots,a_{N}\}$ and imaginary number by $\J$. 
	
	In what follows, we derive PDF and {\color{black}cumulative distribution function} (CDF) of the sum of the product of dGG random variables (in Theorem 1) to develop statistical results of the end-to-end SNR of the considered system (in Theorem 2).

	\begin{my_theorem}
		If $\lvert h_{i,1} \rvert$, $\lvert h_{i,2} \rvert$ are i.ni.d random variables and distributed according to \eqref{eq:dgg_pdf} and $h_{i} = \lvert h_{i,1} \rvert \lvert h_{i,2} \rvert$, then the  PDF and CDF of $h_{\rm RIS} = \sum_{i=1}^N h_i$ are given as
		\begin{eqnarray}\label{eq:pdf1}
			&f_{h_{\rm RIS}}(z) = z^{-1} \prod_{i=1}^{N} A_i B_{i}^{\beta_{i,2}}
			H_{0,1:1,4;...;1,4}^{0,0:4,1;...;4,1} \nonumber \\ &
			\left[\begin{array}{c} U(z) \end{array} \middle\vert \begin{array}{c} - : V_1\\ (1;\{\alpha_{j,2}\}_{j=1}^{N}) : V_2 \end{array} \right] \\
			&F_{h_{\rm RIS}}(z) =  \prod_{i=1}^{N} A_i B_{i}^{\beta_{i,2}}
			H_{0,1:1,4;...;1,4}^{0,0:4,1;...;4,1} \nonumber \\ &
			\left[\begin{array}{c} U(z) \end{array} \middle\vert \begin{array}{c} - : V_1\\ (0;\{\alpha_{j,2}\}_{j=1}^{N}) : V_2 \end{array} \right] \label{eq:cdf1}
		\end{eqnarray}
		where $U(z) = \{\frac{z^{\alpha_{j,2}}}{B_{j}}\}_{j=1}^{N}$, $V_1 = \{{(1,\alpha_{j,2})}\}_{j=1}^{N}$, $V_2 = \{(\beta_{i,2},1),\{(\beta_{i,j},\frac{\alpha_{i,2}}{\alpha_{i,j}})\}_{j=1,3,4}\}_{i=1}^{N}$, $A_{i} = {\frac{\psi_{i,1} \psi_{i,2}}{\alpha_{i,4}} (\phi_{i,2})^{\frac{\alpha_{i,2}\beta_{i,2}-\alpha_{i,4}\beta_{i,4}}{\alpha_{i,4}}}}$ and $B_{i} = {\phi_{i,1}^{-1} (\phi_{i,2})^{-\frac{\alpha_{i,2}}{\alpha_{i,4}}}}$.
	\end{my_theorem}
	\begin{IEEEproof}
		See Appendix A.
	\end{IEEEproof}
	
{\color{black}	Applying maximal ratio combining (MRC) at the destination for the received signals from RIS and direct transmissions},  an expression for the  resultant SNR  is given as $\gamma=\gamma_{RIS}+\gamma_{d} = \gamma_{0, RIS} h_{\rm RIS}^{2} + \gamma_{0,d} \lvert h_{d} \rvert^{2}$,	where $\gamma_{0,RIS}=\frac{h_{l_{ris}}^{2}P_{t}}{\sigma_{v}^{2}}$ and $\gamma_{0,d}=\frac{h_{l}^{2} P_{t}}{\sigma_{v}^{2}}$ represents average SNR for RIS and direct transmissions, respectively.
	
	\begin{my_theorem}
		The  PDF and CDF of the resultant SNR $\gamma=\gamma_{RIS}+\gamma_{d}$ are given as
		\begin{eqnarray}\label{eq:pdf_snr_final}
			&f_{\gamma}(\gamma) = \frac{1}{4} \psi_d \phi_{d}^{-\beta_{d,2}} \gamma^{-1} \prod_{i=1}^{N} A_i B_i^{\beta_{i,2}}
			H_{1,2:1,4;...;1,4;1,2}^{0,1:4,1;...;4,1;2,1} \nonumber \\ &
			\left[\begin{array}{c} U(\gamma) \end{array} \middle\vert \begin{array}{c} (1;\{\frac{\alpha_{j,2}}{2}\}_{j=1}^{N},0) : V_1 \\ (1;\{\frac{\alpha_{j,2}}{2}\}_{j=1}^{N},\frac{\alpha_{d,2}}{2}),(1;\{\alpha_{j,2}\}_{j=1}^{N},0) : V_2 \end{array} \right]	
		\end{eqnarray}	
		\begin{eqnarray}\label{eq:cdf_snr_final}
			&F_{\gamma}(\gamma) = \frac{1}{4} \psi_d \phi_{d}^{-\beta_{d,2}} \prod_{i=1}^{N} A_i B_i^{\beta_{i,2}}
			H_{1,2:1,4;...;1,4;1,2}^{0,1:4,1;...;4,1;2,1} \nonumber \\ &
			\left[\begin{array}{c} U(\gamma) \end{array} \middle\vert \begin{array}{c} (1;\{\frac{\alpha_{j,2}}{2}\}_{j=1}^{N},0) : V_1 \\ (0;\{\frac{\alpha_{j,2}}{2}\}_{j=1}^{N},\frac{\alpha_{d,2}}{2}),(1;\{\alpha_{j,2}\}_{j=1}^{N},0) : V_2 \end{array} \right]	
		\end{eqnarray}
		where $U(\gamma) = \{\{\frac{1}{B_{j}}(\sqrt{\frac{\gamma}{\gamma_{0,RIS}}})^{\alpha_{j,2}}\}_{j=1}^{N},(\phi_d (\sqrt{\frac{\gamma}{\gamma_{0,d}}})^{\alpha_{d,2}})\}$, $V_1 = \{\{(1,\alpha_{j,2})\}_{j=1}^{N};(1,\frac{\alpha_{d,2}}{2})\}$, $V_2 = \{\{(\beta_{i,2},1), \{(\beta_{i,j},\frac{\alpha_{i,2}}{\alpha_{i,j}})\}_{j=1,3,4}\}_{i=1}^{N}; (\beta_{d,2},1),
		(\beta_{d,1},\frac{\alpha_{d,2}}{\alpha_{d,1}})\}$.		
	\end{my_theorem}
	\begin{IEEEproof}
		See Appendix B.
	\end{IEEEproof}
	In what follows, we use the statistical results of Theorem 2 to analyze the system performance using outage probability and average BER.
	\subsection{Outage Probability}\label{sec:outage_probability}
	The outage probability $P_{\rm out}$ is defined as the probability of SNR failing to reach a threshold value, $\gamma_{\rm th}$.  An exact expression for the outage probability is given as  $P_{\rm out}=Pr(\gamma \le \gamma_{\rm th}) = F_{\gamma}(\gamma_{\rm th})$.  We use \cite[eq. 30]{Rahama2018} to express outage probability asymptotically at high SNR as
	\begin{eqnarray}\label{eq:out_high}
		&P_{\rm out}^{\infty} = \frac{1}{4} \psi_d \phi_{d}^{-\beta_{d,2}} \prod_{i=1}^{N} A_i B_{i}^{\beta_{d,2}}  \prod_{i=1}^{N} (\frac{\gamma_{th}}{\gamma_0})^{p_{i}/2}  \nonumber \\  & \Gamma(\beta_{i,1}-\frac{p_{i}}{\alpha_{i,1}}) \Gamma(\beta_{i,2}-\frac{p_{i}}{\alpha_{i,2}}) \Gamma(\beta_{i,3}-\frac{p_{i}}{\alpha_{i,3}}) \Gamma(\beta_{i,4}-\frac{p_{i}}{\alpha_{i,4}}) \nonumber \\ &\Gamma(p_{i}\alpha_{i,2}) (\frac{\gamma_{th}}{\gamma_0})^{p_{d}/2} \Gamma(\beta_{d,1}-\frac{p_{d}}{\alpha_{d,1}}) \Gamma(\beta_{d,2}-\frac{p_{d}}{\alpha_{d,2}})\nonumber\\ & \frac{\Gamma(p_{d}\alpha_{d,2})}{\Gamma(1+\sum_{i=1}^{N}p_{i}\alpha_{i,2}+p_{d}\alpha_{d,2})} \frac{\Gamma(\sum_{i=1}^{N}\frac{p_{i}\alpha_{i,2}}{2})}{\Gamma(\sum_{i=1}^{N}p_{i}\alpha_{i,2})}
	\end{eqnarray}
where $p_{i} = \min\{\alpha_{i,1}\beta_{i,1},\alpha_{i,2}\beta_{i,2},\alpha_{i,3}\beta_{i,3},\alpha_{i,4}\beta_{i,4}\}$ and $p_{d}=\min\{\alpha_{d,1}\beta_{d,1},\alpha_{d,2}\beta_{d,2}\}$. The outage diversity $G_{d\_out}$ is obtained by expressing $P_{\rm out}^{\infty}\propto \gamma_{0}^{-G_{d\_out}}$ to get $G_{d\_out}=\sum_{i=1}^{N} \min\{\frac{\alpha_{i,1}\beta_{i,1}}{2},\frac{\alpha_{i,2}\beta_{i,2}}{2},\frac{\alpha_{i,3}\beta_{i,3}}{2},\frac{\alpha_{i,4}\beta_{i,4}}{2}\}+\min\{\frac{\alpha_{d,1}\beta_{d,1}}{2},\frac{\alpha_{d,2}\beta_{d,2}}{2}\}$. It is clear that the  diversity order depends on RIS elements and  channel fading parameters of reflected and direct transmission links.
	
	\subsection{Average BER}\label{sec:ber}
Using the CDF, the  average  BER for various modulation schemes (parameterized through $a$ and $b$) is given as
	\begin{align}\label{eq:ber}
		\bar{P}_e =&  a \sqrt{\frac{b}{4\pi}} \int_{0}^{\infty} e^{-b\gamma} \gamma^{-\frac{1}{2}} F_{\gamma} (\gamma) d\gamma
	\end{align}

		We substitute $F_{\gamma}(\gamma)$ from \eqref{eq:cdf_snr_final} in \eqref{eq:ber}, use the definition of  multivariate Fox's H-function, and  interchange the order of integration to get
		\begin{flalign}\label{eq:ber2}
			&\bar{P}_e = \big(a\sqrt[]{\frac{b}{4 \pi}}\big) \frac{1}{4} \psi_d \phi_{d}^{-\beta_{d,2}} \prod_{i=1}^{N} A_i B_i^{\beta_{i,2}} \frac{1}{2 \pi j} \int_{L_i} B_{i}^{-s_{i}} \gamma_{0,RIS}^{-\sum_{j=1}^{N}\frac{\alpha_{j,2}s_{j}}{2}} \nonumber \\ &\Gamma(\beta_{i,1}-\frac{\alpha_{i,2}}{\alpha_{i,1}}s_i)  \Gamma(\beta_{i,3}-\frac{\alpha_{i,2}}{\alpha_{i,3}}s_{i}) \Gamma(\beta_{i,4}-\frac{\alpha_{i,2}}{\alpha_{i,4}}s_{i}) \Gamma(\beta_{i,2}-s_i)\nonumber \\ &\Gamma(\alpha_{i,2}s_{i})
			\frac{1}{2 \pi j} \int_{L_d} \phi_d^{s_d} \gamma_{0,d}^{-\frac{\alpha_{d,2}s_{d}}{2}} \Gamma(\frac{\alpha_{d,2}s_d}{2}) \Gamma(\beta_{d,1}-\frac{\alpha_{d,2}}{\alpha_{d,1}}s_d) \nonumber \\ &    
			\frac{\Gamma(\beta_{d,2}-s_d)\Gamma(\alpha-\frac{1}{2}-\frac{\alpha_{d,2} s_{d}}{2})}{\Gamma(\frac{1}{2}+ \alpha)\Gamma(2\alpha-1-\alpha_{d,2} s_{d})} \big(\int_{0}^{\infty} e^{-b\gamma}  \gamma^{\alpha-1} d{\gamma} \big) ds_{i} ds_d
		\end{flalign}
		where $\alpha = \frac{1+\alpha_{d,2}s_d+\sum_{i=1}^{N}\alpha_{i,2}s_i}{2}$.
		We solve the integral as $\int_{0}^{\infty} e^{-b\gamma}  \gamma^{\alpha-1} d{\gamma} = \big(\frac{1}{b}\big)^{\alpha} \Gamma(\alpha)$	and use the definition of $N$-multivariate Fox's H-function \cite[A.1]{M-Foxh} to get eq. \eqref{eq:BER},	as shown at the top of the next page, where			
    	\begin{figure*}[!htbp]
    	\begin{eqnarray}\label{eq:BER}
			&\bar{P}_e = \big(a \sqrt[]{\frac{1}{4 \pi}}\big) \frac{1}{4} \psi_d \phi_{d}^{-\beta_{d,2}} \prod_{i=1}^{N} A_i B_i^{\beta_{i,2}}
			H_{2,2:1,4;...;1,4;1,2}^{0,2:4,1;...;4,1;2,1} 
			\left[\begin{array}{c} U(\gamma_{0,RIS},\gamma_{0,d}) \end{array} \middle\vert \begin{array}{c} (\frac{1}{2};\{\frac{\alpha_{j,2}}{2}\}_{j=1}^{N},\frac{\alpha_{d,2}}{2}),(1;\{\frac{\alpha_{j,2}}{2}\}_{j=1}^{N},0) : V_1 \\ (0;\{\frac{\alpha_{j,2}}{2}\}_{j=1}^{N},\frac{\alpha_{d,2}}{2}),(1;\{\alpha_{j,2}\}_{j=1}^{N},0) : V_2 \end{array} \right]	
		\end{eqnarray}	
		\hrule
		\end{figure*}
$U(\gamma_{0,RIS},\gamma_{0,d}) = \{\{B_{j}^{-1}(\gamma_{0,RIS} b)^{\frac{-\alpha_{j,2}}{2}}\}_{j=1}^{N}, \phi_d (\gamma_{0,d} b)^{\frac{-\alpha_{d,2}}{2}}\}$, $V_1 = \{{(1,\alpha_{j,2})}_{j=1}^{N},(1,\alpha_{d,2})\}$, $V_2 = \{\{(\beta_{i,2},1),\{(\beta_{i,j},\frac{\alpha_{i,2}}{\alpha_{i,j}})\}_{j=1,3,4}\}_{i=1}^{N};(\beta_{d,2},1),(\beta_{d,1},\frac{\alpha_{d,2}}{\alpha_{d,1}})\}$.
	
Similar to outage probability, we can use \cite[eq. 31]{Rahama2018} to express average BER asymptotically at high SNR to derive the diversity of the system as $G_{d\_ber}=\sum_{i=1}^{N} \min\{\frac{\alpha_{i,1}\beta_{i,1}-1}{2},\frac{\alpha_{i,2}\beta_{i,2}-1}{2},\frac{\alpha_{i,3}\beta_{i,3}-1}{2},\frac{\alpha_{i,4}\beta_{i,4}-1}{2}\}+\min\{\frac{\alpha_{d,1}\beta_{d,1}-1}{2},\frac{\alpha_{d,2}\beta_{d,2}-1}{2}\}$.
	
Note that ergodic capacity of the proposed system can be similarly derived and has been left for longer version of the paper.
		\begin{table}[h]
		\caption{dGG fading parameters}
		\label{table:dgg_param}
		\centering
		\begin{tabular}{c c}
			\hline
			Scenario & RIS:\{($\alpha_{1},\beta_{1}$),($\alpha_{2},\beta_{2}$)\}, DT:\{($\alpha_{d,1},\beta_{d,1}$),($\alpha_{d,2},\beta_{d,2}$)\} \\ \hhline{= =} 
			FP1 & \{(2,1),(2,2)\}, \{(1.5,1.5), (1,1.5)\} \\ \hline
			FP2 & \{(1,1),(1,2)\}, \{(2,1.5), (2,1.5)\} \\ \hline
			FP3 & \{(1,1.5),(1,2.5)\}, \{(2,2.1), (2,2.1)\} \\ \hline\hline 
		\end{tabular}
	\end{table}
	\section{Simulation and Numerical Results}\label{sec:sim_results}
	In this section, we validate the derived analytical expressions through Monte-Carlo simulations (averaged over $10^8$ channel realizations) and numerical results.  We consider  carrier frequency $f=6$\mbox{GHz}, antenna gains $G_{T}=10$\mbox{dBi}, $G_{R}=0$\mbox{dBi}, distance from source to RIS, $d_1 = 50$\mbox{m} and RIS to destination,  $d_2=100$\mbox{m}. A noise floor of $-74$\mbox{dBm} is considered over a $20$\mbox{MHz} channel bandwidth. We assume identical dGG parameters for both the hops involving RIS and for each RIS element i.e., $\alpha_{i,1}=\alpha_{i,3}=\alpha_{1}$ and $\alpha_{i,2}=\alpha_{i,4}=\alpha_{2}$ $\forall i$. We use 3 different sets of dGG fading parameters (FP) summarized in table \ref{table:dgg_param}  with $\Omega_{1}=1.5793,\Omega_{2}=0.9671$. For numerical analysis, we used Python code implementation of multivariable Fox's H-function \cite{Alhennawi2016}.

	\begin{figure*}
		\centering
		\subfigure[Outage probability.]{\includegraphics[scale=0.4]{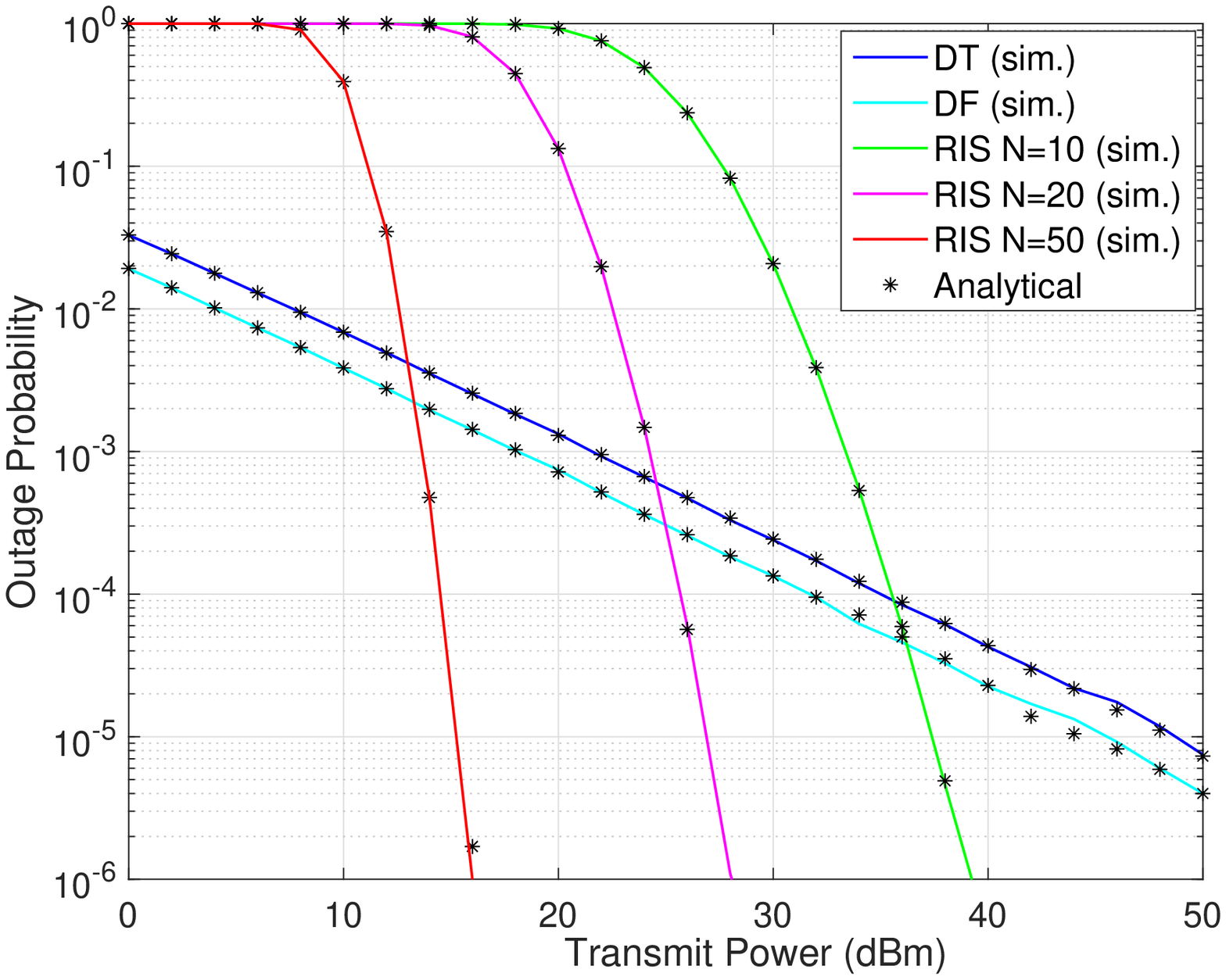}} 
		\subfigure[Average BER.]{\includegraphics[scale=0.4]{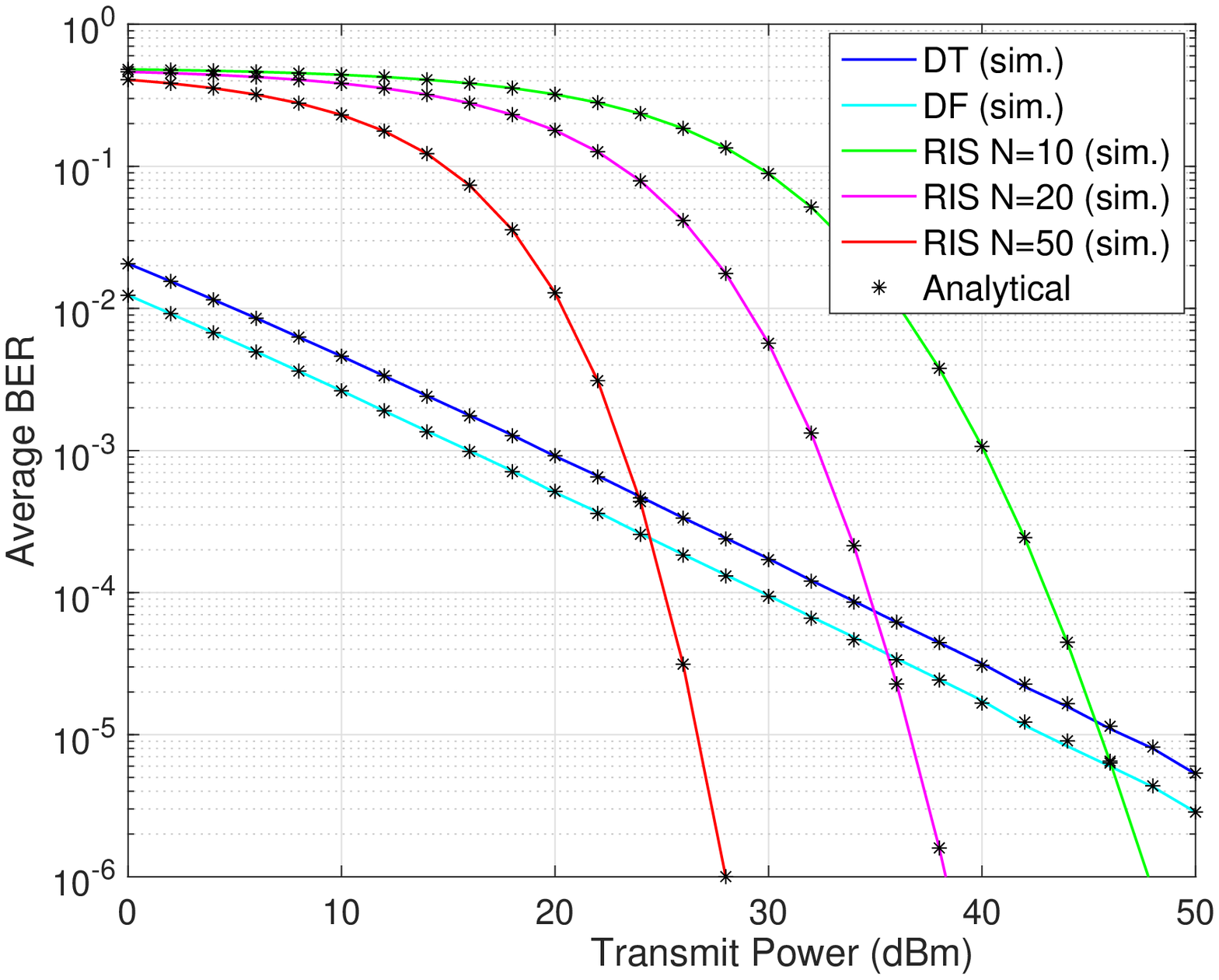}} 
		\caption{Performance scaling of  the RIS-assisted link compared with the DT and relaying systems.}
		\label{fig:outage}
		\label{fig:ber}
	\end{figure*}
	
We demonstrate the performance scaling of RIS-assisted link (without considering signal from direct transmission (DT)) with number of RIS elements to achieve the performance of   stable direct transmission (if available) and decode-and-forward (DF) relaying system for FP1 scenario, as shown in Fig.~\ref{fig:outage}. The plots in Fig.~\ref{fig:outage}(a) and Fig.~\ref{fig:outage}(b) clearly indicate the number of RIS elements required to achieve DT and relaying performance for a given transmit power.   Fig.~\ref{fig:outage}(a) shows that  the outage performance at a lower transmit power is not better due to the multiplicative effect of path loss and short-term fading of the RIS-assisted transmission when compared with DT and relaying systems. However, with an increase in transmit power, the RIS performance surpasses  the DT and relaying system for a given $N$. The figure shows the direct link performance is better  at low transmit powers, and the RIS-system enhances the reliability of the overall system at a higher transmit power. It can be seen that $N=50$ RIS elements are required to achieve DT performance at $P_{t}=15$\mbox{dBm} and just $N=20$ RIS elements at transmit power of $25$\mbox{dBm}. Similarly, Fig.~\ref{fig:outage}(b) shows the average BER performance of the RIS-system for a {\color{black}DPBSK} modulation scheme  with $a=1$ and $b=1$ such that the RIS performance becomes better than the DT and relaying for a given transmit power for a large RIS. Further, slope of the plots in Fig.~\ref{fig:outage}(a) and Fig.~\ref{fig:outage}(b) verify the diversity order of the system.
	
\begin{figure*}
		\centering
		\subfigure[Outage probability.]{\includegraphics[scale=0.4]{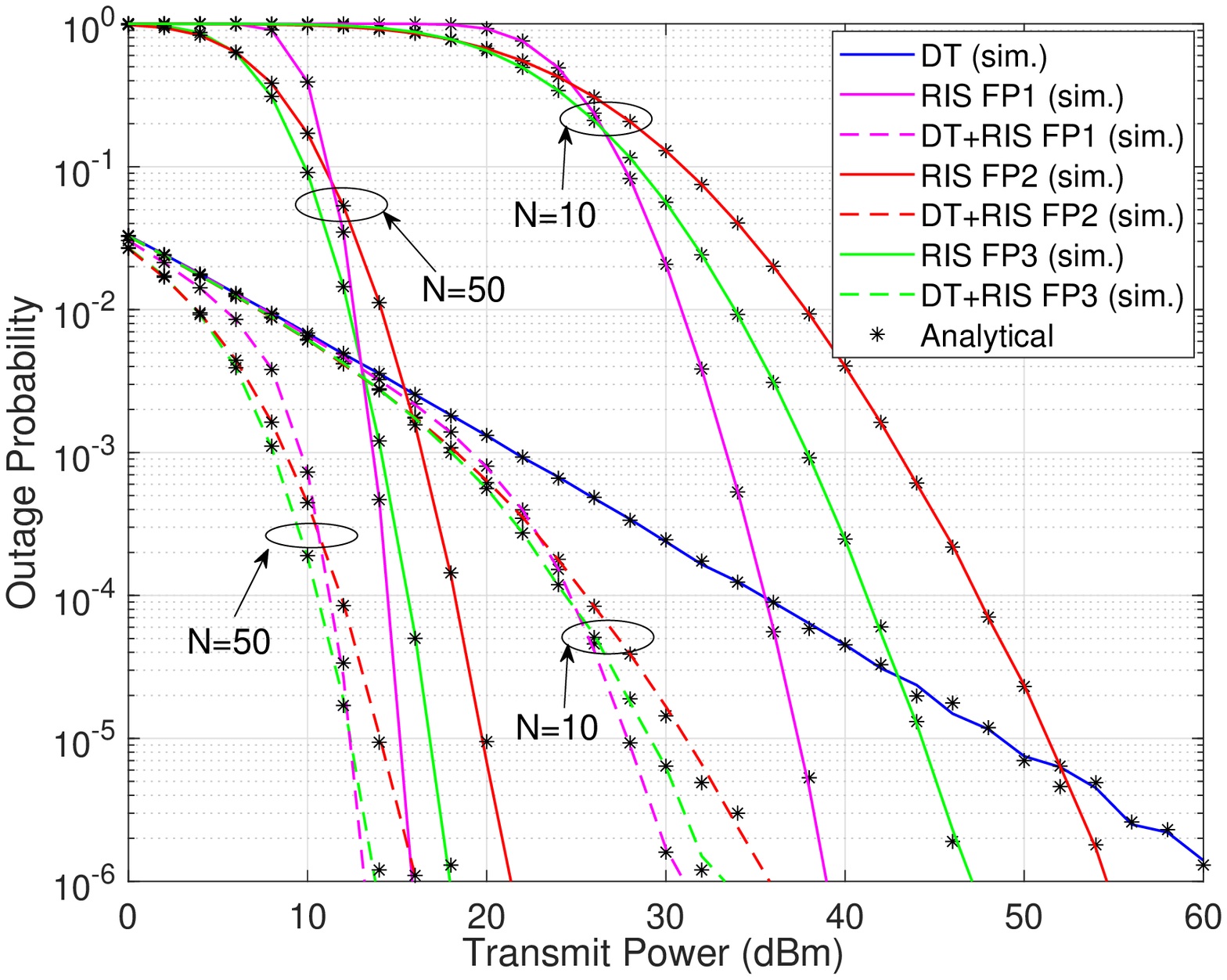}} 
		\subfigure[Average BER.]{\includegraphics[scale=0.4]{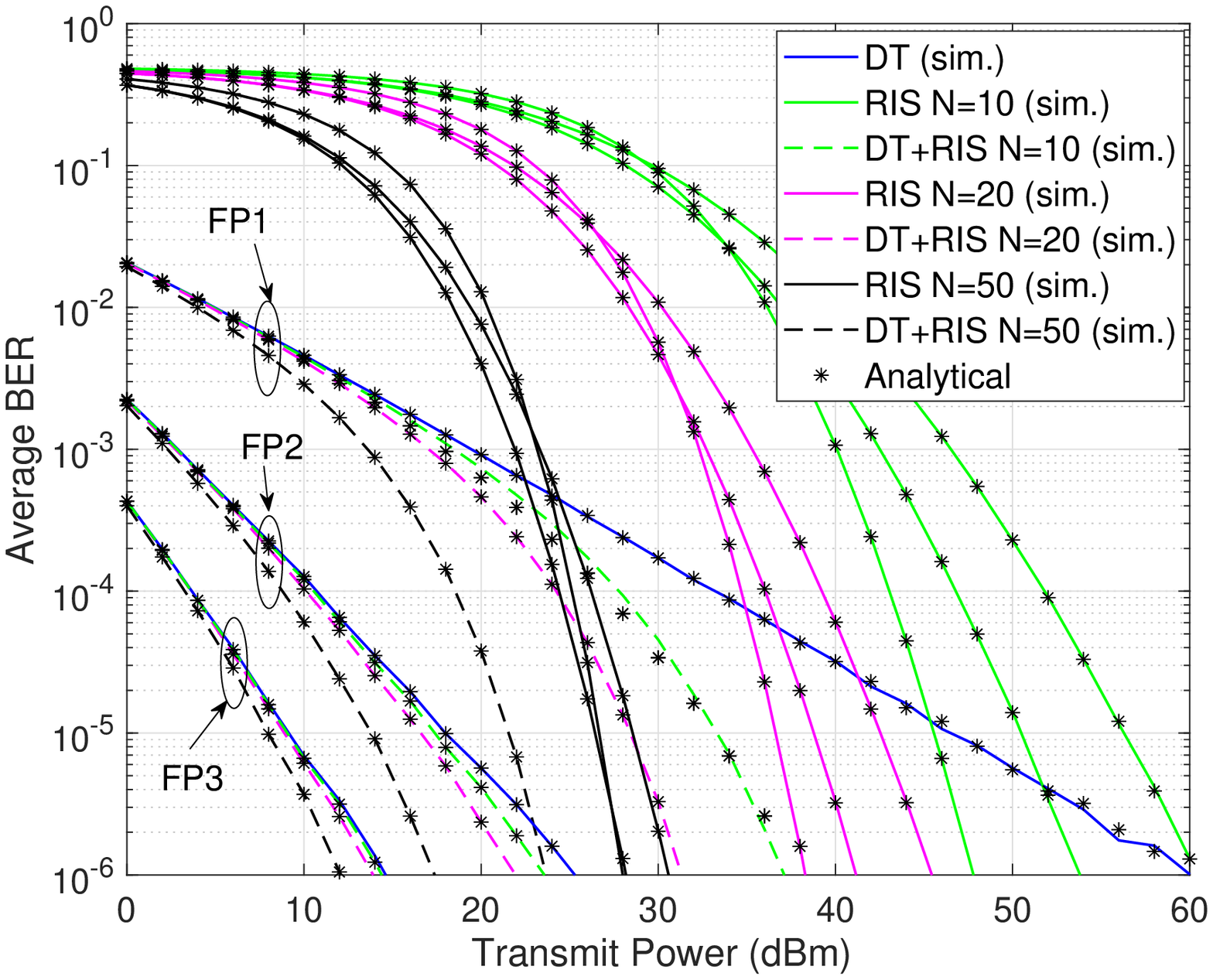}} 
		\caption{Performance of RIS-assisted system with direct transmissions.}
		\label{fig:outage_RIS}
		\label{fig:ber_RIS}
	\end{figure*}

Finally, 	we demonstrate  the performance of RIS-assisted vehicular communications supplemented by the signal from  direct  transmissions, as depicted in Fig.~\ref{fig:outage_RIS}. We consider a higher channel fading scenario (FP1) for the  direct link  and three different fading scenarios (FP1, FP2, and FP3) for the  RIS-assisted link.   It can be seen that the combined effect of RIS and  direct transmission significantly improves the performance at a high transmit power and with a large RIS. As such, for $N=10$ and transmit power less than $15$ \mbox{dBm}, the performance is dictated by the direct link. However,  there is a gain of $15$ \mbox{dBm} to achieve an outage probability of $10^{-4}$ when $N$ is increased from $10$ to $50$.  Comparing FP2 and FP3 scenarios, the performance improves with the increase in shape parameter due to reduction in fading severity: a saving of $5$\mbox{dBm} transmit power when $\beta_{1}$ is changed from $1$ to $1.5$ and $\beta_{2}$ from $2$ to $2.5$. Fig.~\ref{fig:outage_RIS}(b) also demonstrates that the BER performance of the combined system is significantly better than the direct transmission at a high transmit power by harnessing the line-of-sight signal  from the RIS elements. Moreover, the performance of RIS combined with direct transmission is always better than RIS alone (without DT) even at low SNR. Hence, it is clear that combined system performance is better than individual systems  thereby exploiting the presence of DT at low SNRs and the signal received through RIS at high SNRs. It can be seen that the diversity order depends on fading parameters and RIS-elements $N$.

	\section{Conclusion}\label{sec:conc}
	In this paper, we analyzed the performance of a RIS-assisted wireless system that coherently combines the received signals from direct transmission and reflected signals from the RIS module. We derived exact closed-form expressions of PDF and CDF of the end-to-end  SNR considering dGG fading channel models. We presented outage and average BER performance of the considered system in terms of multivariate Fox's H-function. We also derived simplified expressions at high SNR in terms of gamma functions to compute the diversity order of the system.  We used computer simulations to deduce the scaling of RIS-assisted performance compared with the direct transmissions and the conventional relaying system. Further, harnessing the signal from direct link improves significantly the performance of RIS-assisted vehicular network.  Our analysis demonstrates the effectiveness of the coherent combining  of reflected signals from RIS  and the signal from direct link  to achieve reliable performance for a wide SNR range compared with individual systems.
	
	It would be interesting to analyze the system performance considering a mobility model for the vehicular network. Further, the impact of correlated channels and imperfect phase compensation can be investigated as a future scope of the proposed work.

	\section*{Appendix A: PDF and CDF of $h_{\rm RIS}$}
Using $f_{h_{i}}(z) = \int\limits_{0}^{\infty} f_{\lvert h_{i,1} \rvert}(u)f_{\lvert h_{i,2} \rvert}\Big(\frac{z}{u}\Big)\frac{1}{u}du$ with identity  [\cite{M-Foxh} ,2.3], we get the PDF of product of two random variables $h_{i}=\lvert h_{i,1} \rvert \lvert h_{i,2} \rvert$ as
	\begin{eqnarray}\label{eq:pdf_prod}
		&f_{h_{i}}(z) = A_{i} z^{\alpha_{i,2}\beta_{i,2}-1} H_{0,4}^{4,0} \bigg[\begin{array}{c}  B_{i}^{-1} z^{\alpha_{i,2}} \end{array} \big\vert \begin{array}{c} - \\ V_1 \end{array}  \bigg]
	\end{eqnarray}
Next, we use the definition of  moment generating function (MGF) $M_{h_i}(s) =\int_{0}^{\infty}e^{-sz} f_{h_{i}}(z) dz$ and use the identity [\cite{M-Foxh},2.3] to get
	\begin{eqnarray}
	&M_{h_i}(s) = A_i s^{-\alpha_{i,2}\beta_{i,2}} \nonumber \\ &H_{1,4}^{4,1} \bigg[\begin{array}{c}  B_{i}^{-1} s^{-\alpha_{i,2}} \end{array} \big\vert \begin{array}{c} (1-\alpha_{i,2}\beta_{i,2},\alpha_{i,2}) \\ V_1 \end{array}  \bigg] \label{eq:MGF_prod}
\end{eqnarray}
Using  the product of MGF functions with the inverse Laplace transform, we get  the PDF of $h_{\rm RIS}=\sum_{i=1}^N h_i$ as
	$f_{h_{\rm RIS}}(z) = \mathcal{L}^{-1} \prod_{i=1}^{N} M_{h_i}(s)	$. Thus, we use \eqref{eq:MGF_prod}, expand the definition of Fox's H-function, and interchange the order of integration to get
	\begin{align}\label{eq:pdf_appendix}
		&f_{h_{\rm RIS}}(z) = \prod_{i=1}^{N} A_i \frac{1}{2 \pi j} \int_{L_i} B_{i}^{-s_{i}}
		\Gamma(\beta_{i,1}-\frac{\alpha_{i,2}}{\alpha_{i,1}}\beta_{i,2}-\frac{\alpha_{i,2}}{\alpha_{i,1}}s_i)\nonumber \\ & \Gamma(-s_i)\Gamma(\beta_{i,4}-\frac{\alpha_{i,2}}{\alpha_{i,4}}\beta_{i,2}-\frac{\alpha_{i,2}}{\alpha_{i,4}}s_{i}) \Gamma(\beta_{i,3}-\frac{\alpha_{i,2}}{\alpha_{i,3}}\beta_{i,2}-\frac{\alpha_{i,2}}{\alpha_{i,3}}s_{i})
		\nonumber \\ & \Gamma(\alpha_{i,2}\beta_{i,2}+\alpha_{i,2}s_{i}) \Big(\frac{1}{2 \pi j} \int_{L'} s^{\sum_{i=1}^{N}-\alpha_{i,2}\beta_{i,2}-\alpha_{i,2} s_{i}} e^{sz} dt\Big) ds_{i}
	\end{align}

	We apply \cite[8.315.1]{integrals} to solve the inner integral in \eqref{eq:pdf_appendix} as
	\begin{equation}\label{inner_eq}
		\int_{L'} s^{\sum_{i=1}^{N}-\alpha_{i,2}\beta_{i,2}-\alpha_{i,2} s_{i}} e^{sz}ds = \frac{2 \pi j z^{\sum_{i=1}^{N}\alpha_{i,2}\beta_{i,2}+\alpha_{i,2} s_{i}-1}}{\Gamma(\sum_{i=1}^{N}\alpha_{i,2}\beta_{i,2}+\alpha_{i,2} s_{i})}
	\end{equation}

	We use \eqref{inner_eq} in \eqref{eq:pdf_appendix} and apply  the definition of $N$-multivariate Fox's H-function in \cite[A.1]{M-Foxh} to get \eqref{eq:pdf1}. We use similar steps to compute the CDF as $F_{h_{\rm RIS}}(z) = \mathcal{L}^{-1} \frac{\prod_{i=1}^{N} M_{h_i}(s)}{s}$ to get \eqref{eq:cdf1}.
	
	\section*{Appendix B: PDF and CDF of $\gamma=\gamma_{RIS}+\gamma_{d}$}
	To derive the PDF of $\gamma = \gamma_{RIS}+\gamma_{d}$, we first compute the MGF of $\gamma_{RIS}$ and $\gamma_{d}$ and apply the inverse Laplace transform to find the PDF. For MGF of $\gamma_{RIS}$, we use its PDF and expand the definition of multivariate Fox's H-function and interchange the order of integration to get
	\begin{eqnarray}\label{eq:mgf_ris_2}
		&M_{\gamma_{RIS}}(s) =	\frac{1}{2}\prod_{i=1}^{N} A_i B_{i}^{\beta_{i,2}} (\frac{1}{2 \pi \J})^{N} \int_{L_i} B_{i}^{-s_{i}} \gamma_{0,RIS}^{-\frac{\alpha_{i,2}s_{i}}{2}}
		\nonumber \\ & \Gamma(\alpha_{i,2}s_{i}) \Gamma(\beta_{i,1}-\frac{\alpha_{i,2}}{\alpha_{i,1}}s_i) \Gamma(\beta_{i,3}-\frac{\alpha_{i,2}}{\alpha_{i,3}}s_{i})\Gamma(\beta_{i,4}-\frac{\alpha_{i,2}}{\alpha_{i,4}}s_{i})
		\nonumber \\ &\frac{\Gamma(\beta_{i,2}-s_i)}{\Gamma(\sum_{j=1}^{N}\alpha_{j,2}s_{j})}  \Big( \int_{0}^{\infty} \gamma^{\sum_{j=1}^{N}\frac{\alpha_{j,2} s_{j}}{2}-1} e^{-s\gamma} d\gamma\Big) ds_{i}
	\end{eqnarray}

	The inner integral is solved as $\int_{0}^{\infty} \gamma^{\sum_{j=1}^{N}\frac{\alpha_{j,2} s_{j}}{2}-1} e^{-s\gamma} d\gamma =(\frac{1}{s})^{\sum_{j=1}^{N}\frac{\alpha_{j,2} s_{j}}{2}} \Gamma(\sum_{j=1}^{N}\frac{\alpha_{j,2} s_{j}}{2})$.
	
	Similarly, to get MGF of SNR of direct transmission, we express $e^{-s\gamma}$ in terms of Fox's H-function and use the identity [\cite{M-Foxh},2.3] 
	\begin{eqnarray}\label{eq:mgf_dl_3}
		&M_{\gamma_{d}}(s) = \frac{1}{2} \psi_{d} \phi_{d}^{-\beta_{d,2}} \frac{1}{2 \pi \J} \int_{L_d} \phi_{d}^{s_{d}} (s\gamma_{0,d})^{-\frac{s_{d}\alpha_{d,2}}{2}} \nonumber \\ &\Gamma(\beta_{d,2}-s_{d}) \Gamma(\beta_{d,1}-\frac{\alpha_{d,2}}{\alpha_{d,1}}s_d) \Gamma(\frac{\alpha_{d,2} s_{d}}{2}) ds_{d}
	\end{eqnarray}
	
   We solve the inner integral of PDF, $f_{\gamma}(\gamma) = \mathcal{L}^{-1} \big(M_{\gamma_{RIS}}(s) M_{\gamma_{d}}(s)\big)$ as
	\begin{equation}\label{eq:inner_eq_2}
		\int_{L} (s)^{-\sum_{j=1}^{N}\frac{\alpha_{j,2} s_{j}}{2}-\frac{s_{d}\alpha_{d,2}}{2}} e^{s\gamma}  ds = \frac{2 \pi j \gamma^{\sum_{j=1}^{N}\frac{\alpha_{j,2} s_{j}}{2}+\frac{s_{d}\alpha_{d,2}}{2}-1} }{\Gamma(\sum_{j=1}^{N}\frac{\alpha_{j,2} s_{j}}{2}+\frac{s_{d}\alpha_{d,2}}{2})}
	\end{equation}

	Finally, we apply  the definition of $N$-multivariate Fox's H-function in \cite[A.1]{M-Foxh}, to get \eqref{eq:pdf_snr_final}. To compute the CDF of SNR, we use eq. \eqref{eq:pdf_snr_final} in $F_{\gamma}(\gamma) = \int_{0}^{\gamma} f_{\gamma}(x) dx$ and expand $N$-multivariate Fox's H-function in terms of Mellin-Barnes integrals to get
	\begin{eqnarray}\label{eq:cdf_inner_eq_1}
		&F_{\gamma}(\gamma) = \frac{1}{4}\psi_{d} \phi_{d}^{-\beta_{d,2}} \prod_{i=1}^{N} A_i B_{i}^{\beta_{i,2}} (\frac{1}{2 \pi \J})^{N} \int_{L_i} B_{i}^{-s_{i}} \gamma_{0,RIS}^{-\frac{\alpha_{i,2}s_{i}}{2}}
		\nonumber \\ & \Gamma(\alpha_{i,2}s_{i}) \Gamma(\beta_{i,1}-\frac{\alpha_{i,2}}{\alpha_{i,1}}s_i) \Gamma(\beta_{i,3}-\frac{\alpha_{i,2}}{\alpha_{i,3}}s_{i})\Gamma(\beta_{i,4}-\frac{\alpha_{i,2}}{\alpha_{i,4}}s_{i})
		\nonumber \\ &\Gamma(\beta_{i,2}-s_i) \frac{1}{2 \pi \J} \int_{L_d} \phi_{d}^{s_{d}} (s\gamma_{0,d})^{-\frac{s_{d}\alpha_{d,2}}{2}} \Gamma(\beta_{d,2}-s_{d})  \nonumber \\ &\frac{\Gamma(\beta_{d,1}-\frac{\alpha_{d,2}}{\alpha_{d,1}}s_d) \Gamma(\frac{\alpha_{d,2} s_{d}}{2})\Gamma(\sum_{j=1}^{N}\frac{\alpha_{j,2} s_{j}}{2}) }{\Gamma(\sum_{j=1}^{N}\frac{\alpha_{j,2} s_{j}}{2}+\frac{s_{d}\alpha_{d,2}}{2}) \Gamma(\sum_{j=1}^{N}\alpha_{j,2}s_{j})} \nonumber \\ & \Big( \int_{0}^{\infty} x^{\sum_{j=1}^{N}\frac{\alpha_{j,2} s_{j}}{2}+\frac{s_{d}\alpha_{d,2}}{2}-1}  dx\Big) ds_{i} ds_{d} 
	\end{eqnarray} 
	Now, the inner integral can be solved as 
	\begin{eqnarray}\label{eq:cdf_inner_eq_2}
		I=\frac{\Gamma(\sum_{j=1}^{N}\frac{\alpha_{j,2} s_{j}}{2}+\frac{s_{d}\alpha_{d,2}}{2})\gamma^{\sum_{j=1}^{N}\frac{\alpha_{j,2} s_{j}}{2}+\frac{s_{d}\alpha_{d,2}}{2}}}{\Gamma(\sum_{j=1}^{N}\frac{\alpha_{j,2} s_{j}}{2}+\frac{s_{d}\alpha_{d,2}}{2}+1)} 
	\end{eqnarray} 
	We substitute \eqref{eq:cdf_inner_eq_2} in \eqref{eq:cdf_inner_eq_1} and apply the definition of $N$-multivariate Fox's H-function to get \eqref{eq:cdf_snr_final}.
		

	\bibliographystyle{IEEEtran}
	\bibliography{Multi_RISE_full}
	
\end{document}